\documentclass[journal,twoside,web]{ieeecolor}
\usepackage{tmi}
\usepackage{cite}
\usepackage{amsmath,amssymb,amsfonts}
\usepackage{algorithmic}
\usepackage{graphicx}
\usepackage{subcaption}
\usepackage{textcomp}
\usepackage{url}
\usepackage{hyperref}
\usepackage{multirow}
\usepackage{bm}
\usepackage{dirtytalk}
\usepackage{float} % for [H] option
\usepackage{ltablex}
\usepackage{booktabs}
\usepackage{dcolumn}
\usepackage{bbm}
\usepackage{pifont}
\usepackage{xcolor}
\definecolor{hotpink}{HTML}{FF0080}
\definecolor{azureblue}{HTML}{007FFF}

\newcolumntype{P}[1]{D{.}{\pm}{#1}}

\def\BibTeX{{\rm B\kern-.05em{\sc i\kern-.025em b}\kern-.08em
    T\kern-.1667em\lower.7ex\hbox{E}\kern-.125emX}}
\begin{document}
\title{Diff-Def: Diffusion-Generated Deformation Fields for Conditional Atlases}
\author{Sophie Starck, Vasiliki Sideri-Lampretsa, Bernhard Kainz, Martin J. Menten, Tamara T. Mueller, and Daniel Rueckert\\
% \normalsize $^{*}$These authors contributed equally \\
\thanks{S. Starck and V. Sideri-Lampretsa contributed equally to this work. T. T. Mueller and D. Rueckert jointly supervised this work.}
\thanks{S. Starck, V. Sideri-Lampretsa, M. J. Menten, T. T. Mueller and D. Rueckert are with the School of Computation, Information and Technology and the School of Medicine and Health, TUM Klinikum, Technical University of Munich (e-mail: sophie.starck@tum.de, vasiliki.sideri-lampretsa@tum.de, martin.menten@tum.de, tamara.mueller@tum.de, daniel.rueckert@tum.de).}
\thanks{M. J. Menten and D. Rueckert are with the Munich Center for Machine Learning (MCML), Munich, Germany.}
\thanks{B. Kainz, M. J. Menten and D. Rueckert are with the Department of Computing, Imperial College London, UK.}
\thanks{B. Kainz is also with the FAU Erlangen-Nürnberg, Germany (e-mail:bernhard.kainz@fau.de).}
\thanks{This research was conducted using the UK Biobank dataset under the application number 87802. T.T.M., V.S-L and S.S. were supported by the ERC (Deep4MI - 884622). S.S. has furthermore been supported by the German Federal Ministry of Education and Research (BMBF). This research was furthermore supported by the ERC - project MIA-NORMAL 101083647.
The authors gratefully acknowledge the scientific support and HPC resources provided by the Erlangen National High Performance Computing Center (NHR@FAU b143dc, b180dc) of the Friedrich-Alexander-Universität Erlangen-Nürnberg (FAU). The hardware is funded by the German Research Foundation (DFG).}}

\maketitle

\begin{abstract}
Anatomical atlases are widely used for population studies and analysis. 
Conditional atlases target a specific sub-population defined via certain conditions, such as demographics or pathologies, and allow for the investigation of fine-grained anatomical differences like morphological changes associated with ageing or disease. 
Existing approaches use either registration-based methods that are often unable to handle large anatomical variations or generative adversarial models, which are challenging to train since they can suffer from training instabilities. 
Instead of generating atlases directly in as intensities, we propose using latent diffusion models to generate \emph{deformation fields}, which transform a general population atlas into one representing a specific sub-population.
Our approach ensures structural integrity, enhances interpretability and avoids hallucinations that may arise during direct image synthesis by generating this deformation field and regularising it using a neighbourhood of images.
We compare our method to several state-of-the-art atlas generation methods using brain MR images from the UK Biobank. 
Our method generates highly realistic atlases with smooth transformations and high anatomical fidelity, outperforming existing baselines. We demonstrate the quality of these atlases through comprehensive evaluations, including quantitative metrics for anatomical accuracy, perceptual similarity, and qualitative analyses displaying the consistency and realism of the generated atlases.
\end{abstract}


\section{Introduction}
\label{sec:introduction}
\IEEEPARstart{A}{natomical} atlases -- also called templates -- represent the average anatomy of a population in the form of intensity templates or probabilistic maps.
They provide a canonical coordinate system for all images of a cohort and allow for an investigation of inter-subject variability and population differences, as well as anomaly detection \cite{allassonniere2007towards,davis2004large,joshi2004unbiased,bhatia2004consistent,avants2004geodesic,avants2010optimal}. 
An atlas that best represents a whole population should ideally have a minimal morphological distance averaged over all subjects in the dataset. 
However, a single general atlas for the whole cohort is not able to capture the variability between sub-groups, \emph{e.g.}, morphological differences that occur with age. 

As a result, conditional atlases have been introduced to represent a sub-population with specific characteristics (\emph{e.g.}, demographics such as age or sex). Current approaches to create conditional atlases are either based on (a) iteratively aligning images of a sub-group to a reference image or by (b) employing conditional generative models that directly learn the atlas~\cite{dalca2019learning,dey2021generative,li2021cas}.
Usually, the former employs deformable registration \cite{Sotiras2013DeformableMI}, where semantic regions of an image of a cohort and a reference image are aligned and averaged~\cite{Sotiras2013DeformableMI}. 
These methods output a \emph{deformation field} that aligns the image with the atlas, enabling the quantification of anatomical variability and providing insights into structural changes. %, which maps the image to the atlas and can be further used to quantify anatomical variability and interpret the structural changes. 
However, this \say{conventional} approach is time-consuming, as pairwise registration must be recomputed for each condition~\cite{joshi2004unbiased,starck2023constructing,bhatia2004consistent}, and it is highly dependent on the availability of sufficient data. 
Conversely, generative models paired with registration show promising results while being significantly faster \cite{dey2021generative}. 
However, the methods are often greatly affected by training instabilities, hallucinations, and the registration quality, \emph{e.g.}, due to the choice of an inadequate transformation model, potentially leading to low-quality atlases. 

\begin{figure*}[ht!]
    \centering    
    \includegraphics[width=1.0\textwidth]{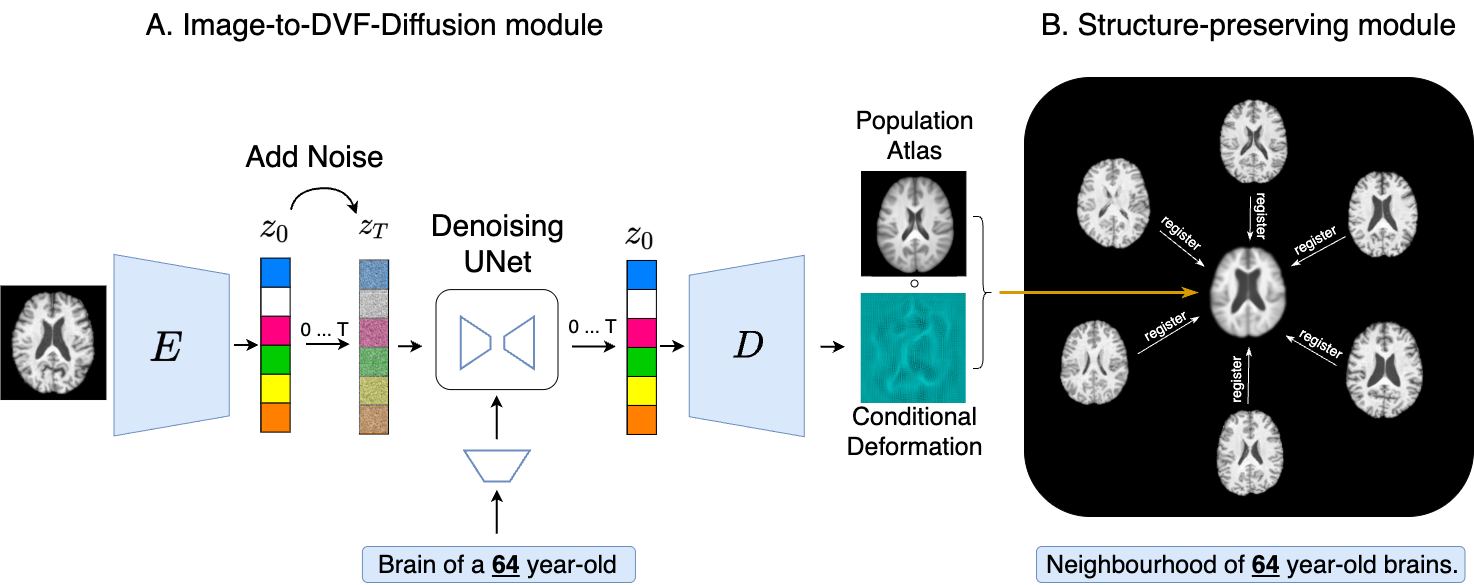}
    \caption{\textbf{Overview of the proposed method.} The latent diffusion module (A), conditioned on a specific attribute of interest, generates a deformation field that warps a general population atlas (\emph{e.g.}, MNI) into a condition-specific atlas. 
    To ensure anatomical plausibility, the morphology-preserving module (B) minimizes the distance between the generated conditional atlas and the subset of training images matching the condition. 
    This encourages the atlas to serve as the most representative sample within its neighbourhood. 
    During inference, the model enables fast and efficient sampling of a conditional atlas given only the target attribute.}
    \label{fig:overview}
\end{figure*}

In this work, we propose to combine the best of both worlds. 
We formulate the task of conditional atlas construction as a \emph{deformation field} generation process using Diffusion Denoising Probabilistic Models (DDPM)~\cite{ho2020denoising}. 
The generated deformation field is used to transform a general population atlas into one representing the sub-group, which is characterised by some desired attributes, \emph{e.g.}, age. To ensure a smooth, anatomically faithful representations, we constrain the conditional atlas to best represent the neighbourhood of images satisfying the attribute of interest. 
Additionally, generating a deformation field enhances the interpretability of the method. Indeed, the deformation field serves as a mapping from a general anatomy to a conditional one, and deformations can be interpreted and quantified as morphological changes. By analysing these deformations, the location and extent of change can be identified, \emph{e.g.},  grey matter atrophy. 
Our core contributions can be summarised as follows: 
 \begin{enumerate}
    \item We utilise diffusion models to generate an interpretable deformation field which transforms a general population atlas into a conditional atlas.
    \item We ensure the construction of a plausible atlas by minimising the distance between the conditional atlas and a representative neighbourhood of images.
    \item We demonstrate the utility of our method by generating brain atlases conditioned on age and ventricular volume and showcase how generating unseen training data results in high-quality atlases.
 \end{enumerate}

\section{Background \& Related Work}

\subsection{Conventional atlas construction}
Anatomical atlases are an important tool in neuroimaging and have been extensively researched for their generation and application in medical image analysis~\cite {paulsen2008detection,thompson2000mathematical}. 
Atlas creation is traditionally performed by iteratively registering all cohort images to a reference image and averaging them \cite{grabner2006symmetric}. 
However, this process is time-consuming and leads to low-quality, blurry atlases that do not capture the details of the underlying structural variability~\cite{avants2010optimal}. 
Furthermore, the selection of a reference image introduces a morphological bias to the appearance of the atlas~\cite{paulsen2008detection,thompson2000mathematical}, requiring an additional unbiasing post-processing step~\cite{joshi2004unbiased} -- further increasing the overall processing time. 
When generating \emph{conditional} atlases with these methods, only a subset of the data is used for each atlas \cite{starck2023constructing}. 
This potentially inhibits the ability to learn features across subsets, and its effectiveness is highly dependent on the decision of the demographic attributes and the availability of relevant data.

\subsection{Learning-based atlas construction}
More recently, generative methods have  become popular for atlas generation. 
They eliminate the data constraint and the additional unbiasing step by learning a conditional atlas without explicitly averaging aligned images~\cite{dalca2019learning,dey2021generative,li2021cas}. 
They are trained with either classic registration objectives~\cite{dalca2019learning}, or generative adversarial networks (GANs)~\cite{dey2021generative}. 
Dalca et al.~\cite{dalca2019learning} propose a network that generates a conditional diffeomorphic (\emph{i.e.} differentiable, invertible, and smooth) atlas. 
However, the diffeomorphic transformation model may be inadequate, resulting in lower quality atlases due to the intricate nature of human anatomy, which is often non-smooth, \emph{e.g.}, when registering healthy to pathological images. To address this, Dey et al.~\cite{dey2021generative} propose a GAN-based model, combined with non-diffeomorphic registration, that simultaneously minimises a registration and an adversarial loss. 
While this shows promising results, GANs are challenging to deploy as they suffer from training instabilities and mode collapse~\cite{arjovsky2017wasserstein,gulrajani2017improved,mescheder2018convergence}. 
For these reasons, in this work, we leverage the capabilities of diffusion models.

\subsection{Diffusion models} 
Recently, diffusion models~\cite{sohl2015deep,ho2020denoising} have emerged as robust probabilistic generative models designed to capture and learn complex data distributions.
More specifically, score-based denoising diffusion probabilistic
models (DDPMs)~\cite{ho2020denoising}, have shown remarkable performance in generative modelling in various computer vision fields~\cite{croitoru2023diffusion,yang2023diffusion,chen2024overview}. 
While they are capable of yielding high-fidelity data, unlike GANs, they also provide attractive properties such as scalability and training tractability~\cite{chen2024overview, kazerouni2023diffusion}. 
In addition, diffusion models have been used in the medical imaging domain for various tasks~\cite{khader2022medical,kazerouni2023diffusion,packhauser2023generation,dorjsembe2022three,chambon2022roentgen,sanchez2022healthy}, such as conditional synthetic image generation~\cite{pinaya2022brain,moghadam2023morphology,dorjsembe2022three, bedel2024dreamr}, anomaly detection~\cite{bercea2023mask,wolleb2022diffusion}, image-to-image translation~\cite{lyu2022conversion,meng2022novel,ozbey2023unsupervised} and registration~\cite{kim2022diffusemorph,kim2022diffusion,qin2023fsdiffreg}.
Specifically in the context of image registration,~\cite{kim2022diffusemorph,kim2022diffusion,qin2023fsdiffreg} utilise the spatial information encoded in the latent feature vectors estimated by diffusion models to generate deformation fields for pairwise image registration. 
However, since these methods focus on registration, an additional step is still required in order to generate an atlas, requiring a sub-population split and aggregation, similar to the conventional atlas generation methods.

\section{Methods}

In this work, we propose a novel approach for learning sub-population-specific atlases by leveraging the generative capabilities of conditional latent diffusion models (LDMs)~\cite{rombach2022high}. 
Specifically, we train an LDM to generate high-resolution 3D deformation vector fields (DVFs) conditioned on a given attribute. 
These generated DVFs enable the transformation of a general population atlas to align with the characteristics of the target sub-population. 
To ensure anatomical fidelity, we introduce a morphology preservation component based on deformable registration, which constrains the generated atlas to maintain biologically plausible structures while adhering to the specified conditioning attributes.
An outline of the proposed method is illustrated in Figure~\ref{fig:overview}).

% These high-quality deformation vector fields can then be used to transform a general population atlas to fit the desired conditioning characteristics. 

% Our proposed method (illustrated in Figure~\ref{fig:overview}) leverages the generative capabilities of conditional latent diffusion models (LDMs)~\cite{rombach2022high} to generate high-resolution 3D \emph{deformation vector fields} (DVFs). 
% These deformation vector fields can then be used to transform a general population atlas to fit the desired conditioning characteristics. 

\subsection{Deformation field synthesis} 
\label{sec:def_synthesis_ldm}

To generate high-quality conditional atlases based on a feature of interest, we employ the capabilities of the Denoising Diffusion Probabilistic Models (DDPM) \cite{ho2020denoising}, which has demonstrated promising results in both natural and medical image synthesis \cite{yang2023diffusion, pinaya2022brain}. 
Specifically, this study aims to synthesise reliable conditional deformation fields \(\phi_{c}:\mathbbm{R}^3\rightarrow\mathbbm{R}^3\), to transform a population atlas, \emph{e.g.} the MNI atlas (\(\mathcal{A}_{\text{MNI}}\))~\cite{grabner2006symmetric} to an atlas that satisfies a certain condition \(c\) (\(\mathcal{A}_{final} = \mathcal{A}_{\text{MNI}} \circ \phi_{c}\)).
However, DDPMs are notorious for their high memory requirements, particularly when handling high-resolution 3D data.
For this reason, to be able to scale to high-resolution deformation fields, we opt for using Latent Diffusion Models (LDM)~\cite{rombach2022high}, which enable diffusion model training on limited computational resources while retaining their quality and flexibility. 

LDMs decompose the generation process into a sequential application of autoencoders (AE) and denoising diffusion models.
A 3D brain image is projected into the latent space during the forward process. 
Then, Gaussian noise \(\mathcal{N}(0,1)\) is iteratively introduced to the latent variable through a fixed Markov chain, gradually degrading its content.
During the reverse process, modelled as a Markov chain, the model learns to recover the signal given the noisy input and a conditional vector based on the attributes of the sub-population of interest, learning, \emph{i.e.}, age and ventricular volume.
The resulting denoised latent variable, which contains spatial and anatomical features from the input image is finally decoded into a high-resolution deformation field.
This is achieved by feeding the denoised latent variable into a convolutional decoder which outputs the deformation field of size \([B, 3, H, W, D]\) where \(B\) denotes the batch size, \(3\) denotes the \(x, y\) and \(z\) components of the deformation field while \(H, W, D\).
This deformation field is subsequently used to warp the general population atlas to a condition-specific atlas.

% \begin{figure}[ht!]
%     \centering    
%     \includegraphics[width=0.9\columnwidth]{figures/tmi_neigh1.png}
%     \caption{Gaussian vs. uniform sampling for neighbourhood selection.
%     Using Gaussian sampling to select neighbours around a target condition ensures that samples closer to the condition of interest are more likely to be selected while still allowing the inclusion of samples from a broader range. 
%     This results in a focused yet diverse neighbourhood promoting specificity and preserving variability. 
%     In contrast, uniform sampling considers all nearby values equally, which may diminish the relevance of the conditioning and reduce the representativeness of the generated atlas.}
%     \label{fig:neighbourhood}
% \end{figure}

The LDM therefore learns to represent data in a structured latent space, ensuring that the generated deformation fields are not random or noisy, but instead follow a process that aligns with the physical and spatial characteristics of the input images.
Additionally, since the diffusion process is performed over multiple steps, the model becomes more adept at generating deformation fields that are not only realistic but also generalisable across different images with similar demographic characteristics. 
Rather than relying on the latent vector of a single image, the model learns to generate meaningful deformation fields that can be applied to any modality.
An illustrative representation of the process is shown in Fig.~\ref{fig:overview} (A).

\subsection{Morphology preservation} 
\label{sec:morphology_preservation_implementation}

The conditional deformation fields \(\phi_c\) generated by the diffusion process described in the previous section~\ref{sec:def_synthesis_ldm}, allow us to flexibly deform a general population atlas.
However, we have to ensure that the resulting condition-specific atlas \(\mathcal{A}_{\text{final}}\) is anatomically faithful and compliant with the demographic feature of interest.
For this reason, we introduce a differentiable \emph{morphology preservation} component based on deformable registration (Figure \ref{fig:overview} (B)) to guarantee that the generated deformation field yields a high-quality condition-specific atlas that preserves the anatomical cues.

\begin{figure*}[ht!]
    \centering
    \includegraphics[width=0.95\textwidth]{figures/csf_growing.png}
    \caption{Overview of the generated atlases conditioned on ventricular volume. 
    Our method generates displacement fields that deform a general population atlas to match specific conditions, enabling precise quantification of spatial changes.
    The first row illustrates the Jacobian determinant of the deformation field $\mathcal{J}(\phi_{c}$).
    Expansion in the image domain is denoted in red, while contraction is in blue.}
    \label{fig:csf}
\end{figure*}

An atlas defines a common reference space for all images and represents an \emph{average} image derived from the whole cohort.
Consequently, a condition-specific atlas should be an \emph{average} representation of a \emph{neighbourhood} of images that satisfy a demographic trait, \emph{e.g.}, a 65-year-old brain.
Based on this intuition that the conditional atlas should minimise the distance to each image in the condition-specific \emph{neighbourhood}, we build the proposed differentiable \emph{morphology preservation} component.
Given a condition $c$, we sample a neighbourhood of N images that satisfy that condition and using deformable image registration~\cite{balakrishnan2019} we obtain a deformation field \(\phi_i\) between each image \(i\) in the neighbourhood N and the condition-specific atlas \(\mathcal{A}_{\text{final}}\).
This field, \(\phi_i\), serves as a voxel-wise measure of structural distance between image pairs, \emph{i.e.}, the displacement of each neighbourhood image relative to the generated atlas.
To ensure that the conditional atlas \(\mathcal{A}_{\text{final}}\) is effectively the \emph{average} representation of this \emph{neighbourhood} that satisfies the condition, we ensure that its geodesic distance to every image in the neighbourhood is minimal.
This is represented in Figure \ref{fig:overview} (B).

\subsection{Training and supervision} 
\label{sec:training}
The proposed approach consists of three distinct components. 
An Autoencoder (AE) is employed to generate a latent representation of the input data, facilitating scalability to high-resolution 3D medical images. 
A Latent Diffusion Model (LDM) is utilised to synthesise the high-resolution deformation field. 
Finally, a morphology preservation (MP) module built upon deformable image registration is integrated to ensure that the high-resolution deformation field maintains anatomical fidelity and complies with the given condition. 
% Finally, a deformable registration network is responsible for aligning the conditional atlas with each image in the surrounding neighbourhood.

\subsubsection{Autoencoder} 
The autoencoder (AE) is pre-trained to learn a compressed latent representation $z$ for each image.
% We denote the encoder and decoder parts of the AE with \(E\) and \(D\), respectively.
Following \cite{pinaya2022brain}, the autoencoder's objective function is a combination of $L_1$ loss between an image \(I\) and its reconstructed pair \(I_{\text{recon}}\), perceptual loss~\cite{zhang2018unreasonable} \(\mathcal{L}_{\text{perc}}\), an adversarial objective \(\mathcal{L}_{\text{adv}}\) operating on patches~\cite{esser2021taming} (\(p, p_{recon}\)) and a KL latent space regulariser \(\mathcal{L}_{\text{KL}}\) as follows:

\begin{equation}
\begin{split}
\mathcal{L}_{\text{AE}} = &\mathcal{L}_{\text{$L_1$}}(I, I_{\text{recon}}) 
+ \lambda_1 \mathcal{L}_{\text{perc}}(I, I_{\text{recon}}) \\
&+ \lambda_2 \mathcal{L}_{\text{adv}}(p, p_{\text{recon}}) \\
& + \lambda_3 \mathcal{L}_{\text{KL}}(q(z|I) \| p(z)).
\end{split}
\end{equation}
Here \(q(z|I)\) is the distribution generated by the encoder \(E_a\) and \(p(z) = \mathcal{N}(0, 1)\). Each \(\lambda\) is a weighting factor for its respective loss.
% This AE provides a latent representation of an image through \(E_a\) and decodes it back to an image with \(D_a\).
% For the following steps, we freeze the encoder \(E_a\) to provide a latent representation of an image, and the decoder \(D_a\) is further optimised to generate DVFs (Section \ref{sec:def_synthesis}).

\subsubsection{Latent Denoising Diffusion Model (LDM)}
Next, we utilise the latent representation previously learned from the autoencoder as an input to train a conditional latent diffusion model to synthesise high-quality deformation fields.
For this reason, while we freeze the encoder \(E\), we keep the decoder \(D\) trainable, changing its last layer's output channels from 1 to 3. This ensures the output to be 3D deformation fields instead of images, \emph{i.e.} deformation vectors instead of intensity scalars.
The desired deformation field \(\phi_c\) conditioned on condition \(c\) is then synthesised by feeding the denoised latent vector $z_0'$ to the decoder \(D\).
Having this pre-trained decoder is a crucial step since it allows us to maintain useful structural cues contained in the image while learning to map those to a deformation field.
% Moreover, starting from the pre-trained decoder and not training it from scratch ensures training stability and faster convergence.

Following~\cite{rombach2022high,pinaya2022brain}, we effectively condition the model using a hybrid approach combining concatenation of the conditioning with the input data and the use of cross-attention mechanisms~\cite{rombach2022high}.
The overall loss function for the LDM can be expressed as:

\begin{equation}\label{eq:diffusion1}
    \mathcal{L}_{\operatorname{\text{diff}}} = \mathbb{E}_{x, \epsilon, t, c} \left[ \|\epsilon - \epsilon_{\theta}(x_t, c, t) \|_2^2 \right]
\end{equation}

Here N is the number of samples, \(\epsilon_i \) is the true noise for the \( i \)-th sample, and \( \epsilon_\theta(z_{t_i}, c_i, t_i) \) is the predicted noise from the model with \( z_{t_i}\) being the latent variable, \(c_i\) the condition, and \(t_i\) the time step.

% Furthermore, we learn the desired projection to the deformation domain that satisfies the query condition, by minimising the morphology preservation loss (Eq.~\ref{eq:eq_morph} in Section~\ref{sec:morphology_preservation}). 

\subsubsection{Morphology preservation} 
\label{sec:morphology_preservation}
The conditional deformation fields generated by the diffusion process allow us to flexibly deform a general population atlas to the targeted, conditional atlas \(\mathcal{A}_{\text{final}}\). 
In order to ensure that the deformed population atlas satisfies the condition of interest while preserving anatomical cues, we introduce a \emph{morphology preservation} component based on deformable registration (Figure \ref{fig:overview} B).

Each image \(i\) of a selected neighbourhood N is aligned to the conditional atlas \(A_{\text{final}}\) using deformable image registration, generating a deformation field \(\phi_i\).

\begin{equation}\label{eq:A_{final}}
A_{\text{final}} = A_{\text{MNI}} \circ \phi_{c} ,
\end{equation}

To reduce the computational time required for deformable registration, we leverage the advantages of learning-based registration, which significantly accelerates pairwise image alignment compared to traditional iterative optimisation methods.
More specifically, we use a UNet-based convolutional registration network~\cite{balakrishnan2019}, which we have pre-trained to perform pairwise registration on our dataset.

% A deformation field, \(\phi_i\), is then obtained by registering each image \(i\) of the neighbourhood N to the conditional atlas, \(A_{\text{MNI}} \circ \phi_{c}\), using 

\begin{equation}\label{eq:phi}
\phi_i = f_{\theta}(\mathcal{A}_{\text{MNI}} \circ \phi_{c},N_i),
\end{equation}

\noindent where \(A_{\text{MNI}}\) is the MNI atlas~\cite{grabner2006symmetric} and \(\phi_{c}\) is the condition-specific deformation field generated by the diffusion process and \(N_i\) denotes the $i^{th}$ data point in the neighbourhood, where $i \in [1, N]$, that satisfies the condition $c$, and \(f_{\theta}\) the pre-trained UNet, with parameters \(\theta\).

% To learn a suitable conditional atlas we optimise the conditional deformation field output by the LDM, \(\phi_c\), minimizing the following loss function:

% \begin{equation}\label{eq:eq_morph}
%     \mathcal{L}_{\text{morph}} = \frac{1}{N}\lVert \phi_i\rVert_2^2.
% \end{equation}

% \noindent where \(\phi_i\) is the average deformation of each neighbouring image to the conditional atlas \(A_{\text{MNI}} \circ \phi_{c}\), \(A_{\text{MNI}}\) is the MNI atlas~\cite{grabner2006symmetric} and \(\phi_{c}\) is the condition-specific deformation field.

To ensure that the conditional atlas accurately represents the average structure of the neighbourhood satisfying the given condition, its distance to each image within this neighbourhood should be minimised.
This can be realised by minimising the following loss function:

\begin{equation}\label{eq:eq_morph}
    \mathcal{L}_{\text{morph}} = \frac{1}{N} \sum_{i}^{N}\lVert \phi_i\rVert_2^2.
\end{equation}

\noindent where \(\phi_i\) is given by Eq.~\ref{eq:phi}

We use Gaussian sampling as a heuristic to favour images that better match the condition, increasing the likelihood of selecting those images over others. 
This sampling is non-deterministic, allowing the sampling of different neighbourhoods at each epoch and, therefore, allowing the model to learn from a larger range of data.
The advantage of Gaussian sampling is that it comprises more samples closer to the condition of interest while still sampling sparser values further away, which is especially valuable for conditions where we have missing data.
% This sampling is displayed in Fig.~\ref{fig:neighbourhood}.

\subsubsection{Overall supervision} 
\label{sec:supervision}

The combination of diffusion and morphology preservation allows us to obtain smooth, stable, and geometrically plausible conditional atlases and guarantees that the atlas reflects the feature of interest. 
This design is also able to learn the underlying data distribution, which has the advantage of interpolating between conditions \say{seen} during training, modelling a continuous data distribution, \emph{e.g.}, ageing.

The whole approach is trained end-to-end to minimise the following loss function:

\begin{equation}\label{eq:eq2}
    \mathcal{L} = 
    \mathcal{L}_{\operatorname{\text{diff}}} + \alpha\mathcal{L}_{\operatorname{\text{morph}}} + \beta\mathcal{R}(\phi_{c}).
\end{equation}

The overall objective is a linear combination of three terms. First, the diffusion loss $\mathcal{L}_{\text{diff}}$, controlling the representation generation.
Secondly, the morphology preserving loss $\mathcal{L}_{\text{morph}}$, enforcing neighborhood similarity.
The third term is a bending energy term~\cite{rueckert1999nonrigid} enforcing smoothness on the conditioned deformation field.
Finally, $\alpha,\beta$ are weighting factors determined experimentally, controlling each component's contribution to the overall objective.

\subsection{Inference} 
\label{sec:inference}
During inference, we feed the diffusion model with a random noise vector and an embedded condition vector, using the hybrid conditioning approach described in \cite{rombach2022high,pinaya2022brain}. The model then performs iterative denoising over 500 time steps. The resulting latent vector is passed through the decoder to produce the deformation field. Finally, this deformation field is applied to warp the general (MNI) atlas, resulting in an atlas that satisfies the condition of interest.

\begin{figure*}[t]
    \centering
    \includegraphics[width=\textwidth]{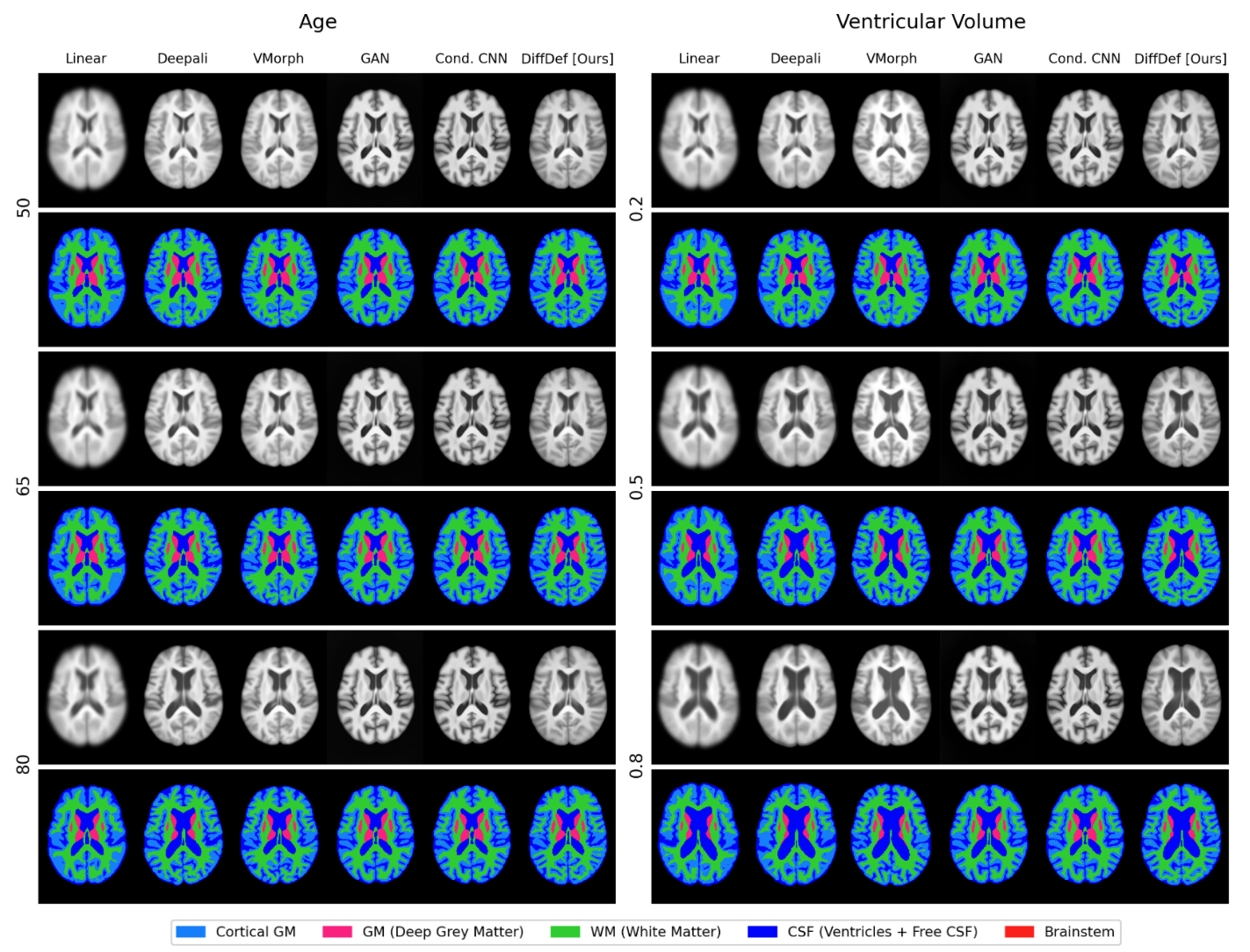}
    \caption{Qualitative results of the proposed method (DiffDef) and baseline models, conditioned on normalized \textbf{ventricular volumes} (\(0.2,0.5,0.8\)) and \textbf{ages} (\(50,65,80\) years).
    DiffDef, the only method that generates displacement fields, effectively captures the anatomical progression associated with both conditions, \emph{e.g.} the growth of the ventricular volume in both cases. 
    At the same time, it preserves the appearance characteristics of the original cohort, maintaining consistency with the underlying intensity distribution.}
    \label{fig:qualitative}
\end{figure*}

\section{Experimental Setup}

\subsection{Dataset} We use $5000$ T1-weighted brain Magnetic Resonance Images (MRI) from the UK Biobank~\cite{sudlow2015uk}. 
More specifically we use $4000$ images for training, $300$ for validation and hyperparameter tuning and $700$ for testing, i.e. conditional atlas creation using the conventional methods. 
The brain images have an isotropic spacing of \(1\rm{mm}^3\) and a size of $160\times225\times160$. 
All images are skull-striped using BET~\cite{smith2000bet}, rigidly registered to a common MNI space~\cite{grabner2006symmetric} using the conventional registration framework Deepali~\cite{deepali}, and segmented using SynthSeg~\cite{billot_synthseg_2023}. The resulting segmentations contain 31 detailed labels of the brain regions that we categorise into four: cortical grey matter, deep grey matter, white matter, cerebrospinal fluid and brainstem.
As conditions, we use the subjects' age, ranging from $50$ to $80$ years old and the ventricular volume normalised by the total number of voxels, ranging from $0.0$ to $0.6$. 
Furthermore, we use the publicly available MNI ICBM152 template~\cite{grabner2006symmetric} as a general population atlas for the brain data.

\begin{table}[!h]
% \small
\addtolength{\tabcolsep}{4pt}
\centering
\caption{Selected hyperparameters for each baseline. We refer the reader to the relevant papers for further details regarding the architectural choices.}
    \resizebox{\columnwidth}{!}{
    \begin{tabular}{lccccc}
    \toprule
    & \textbf{Deepali}~\cite{deepali} & \textbf{GAN}~\cite{dey2021generative} & \textbf{VXM}~\cite{Qiu2021LearningDA}& 
    % \textbf{Img LDM}~\cite{pinaya2023generative} & 
    
    \textbf{Cond. CNN}~\cite{dalca2019learning}\\
    \midrule
    \textbf{Learning Rate} & $10^{-3}$ & $10^{-4}$ & $10^{-4}$& 
    % $2.5^{-5}$ & 
    $10^{-4}$\\
    \textbf{Reularisation \(\lambda\)}& $10^{-1}$ & $10^{-3}$ & $10^{-1}$ 
    % & - 
    & $10^{-3}$\\
    \textbf{Batch Size} & - & $1$ & $8$ 
    % & $1$ 
    & $1$\\
    \textbf{Resolution levels}& $3$ & $1$ & $1$ 
    % & $1$
    & $1$\\
    \bottomrule
    \end{tabular}
}
\label{tab:hyperparameters}
\end{table}

\subsection{Implementation} We implement the AE and the core LDM components following~\cite{rombach2022high,pinaya2022brain} and using the publicly available repository~\cite{pinaya2023generative}.
The AE is trained with a learning rate of $5e^{-5}$, batch size of 1 and embedding size of $20 \times 28 \times 20$, with weighting coefficients set to \(\lambda_1 = 0.002\), \(\lambda_2 = 0.005\), \(\lambda_3 = 10^{-8}\). These coefficients were set experimentally by optimising perceptual similarity between the input image and its reconstructed counterpart. This latent representation is then fed to the diffusion model, which uses a learning rate of $2.5^{-5}$ and a DDPM scheduler, batch size of 1 and noise scheduling with $1000$ steps.

The morphology-preservation controls the atlas plausibility through a neighbourhood loss. 
The amount of samples in the neighbourhood influences the stability of the method, as well as the memory requirements. 
Consequently, a trade-off between performance and memory needed to be met, and the number of neighbours N was set to $15$, which is the maximum number of neighbours that did not exceed memory requirements and yielded the best comparative results (see Table~\ref{tab:ablation_results}).
Moreover, the neighbourhood was sampled following a Gaussian weighting scheme where the probability of selecting each value is influenced by its proximity to a central value. 
This mean corresponds to the morphological features associated with the desired condition, as described in Section \ref{sec:morphology_preservation_implementation}.
This approach ensures that values closer to the mean are given a higher weight, while those further from the mean are increasingly less likely to be selected.
We choose a value of sigma \(\sigma = 0.05\) to limit the selection to values that are very close to the central value, limiting the number of influential neighbours.

Each image of the neighbourhood is then registered to the generated atlas using a pre-trained registration network~\cite{balakrishnan2019} in evaluation mode, leveraging its fast inference times. 
The model is implemented using the U-Net registration model that is publicly available~\cite{qiu2021learning} and is trained with a learning rate of $10^{-4}$ using a batch size of $8$ with cross-correlation~\cite{lewis1995fast} as a distance metric and a regularisation term weighted by \(\lambda_{Reg}=0.1\).
The resulting loss is a linear combination of the diffusion loss, the morphology preserving loss and a regularisation term.
We train all models to converge and retain the optimal hyperparameters based on the validation set.
We specifically obtain a weight of $1$ to the morphology-preserving loss $\alpha$ and $0.5$ to the regularisation term $\beta$. 
More detailed implementation details regarding the AE, LDM, registration network, and baseline hyperparameters can be found in Table~\ref{tab:hyperparameters}. 
We train all networks on an A100 80GB GPU with Pytorch. 
The source code is publicly available\footnote{\url{https://github.com/starcksophie/DiffDef/}}.

\subsection{Baselines} We compare our method to five related approaches for atlas generation.
We evaluate three widely used \emph{unconditional} atlas construction algorithms: a linear average of the images, Deepali~\cite{deepali}, an iterative optimisation registration framework based on the MIRTK software \cite{mirtk}, and Voxelmorph~\cite{balakrishnan2019}, a learning-based method. 
Since these methods are unconditional, we sample, register, and average $1000$ subjects for every condition to generate conditional atlases. Deepali (DLI) and Voxelmorph (VXM) are registration frameworks; an extra step is required to produce the atlas. 
The models are used to register the images to an arbitrary reference image. 
They are subsequently averaged, generating a first atlas biased towards the reference image \cite{joshi2004unbiased}. 
Following existing approaches~\cite{rueckert1999nonrigid, joshi2004unbiased}, the initial atlas is unbiased by averaging the resulting DVFs and applying the corresponding transformation.

Furthermore, we use two \emph{conditional} learning-based methods as baselines. 
We investigate a GAN-based method~\cite{dey2021generative} consisting of a convolutional generator conditioned on the attribute of interest to produce the desired atlases.
These are further registered to every image in the dataset, and then a discriminator ensures that the resulting atlases have a realistic appearance. 
Additionally, we investigate the conditional learning-based convolutional network (Cond. CNN) proposed by Dalca et al.~\cite{dalca2019learning}, which features a convolutional decoder that takes the condition as input and generates a residual, subsequently added to a linear average of all the images in the dataset. 
Similarly to the previous method, the resulting atlases are further registered to every image in the dataset.

\begin{table*}[h]
% \small
\addtolength{\tabcolsep}{2pt}
\centering
\caption{Quantitative results that assess competing properties (anatomical and appearance similarity) of the generated atlases conditioned on the age and ventricular volume. 
We perform pairwise comparisons between each generated atlas and a test set of $100$ images for each condition and report each metric's mean and standard deviation.
The best results are highlighted in \textbf{bold}, and the second best are \underline{underlined}.
}
    \resizebox{\textwidth}{!}{
    \begin{tabular}{lccccc}
    \toprule
    \multicolumn{6}{c}{\textbf{Age}} \\
    \midrule
    & \textbf{DSC} $\uparrow$ & \textbf{Folding (\%)}  $\downarrow$ & \textbf{Smoothness} $\downarrow$ & \textbf{Avg. disp.} $\bm{\lVert \Phi \rVert}$ $\downarrow$ & \textbf{LPIPS} $\downarrow$\\ 
    \midrule
    \textbf{Linear} & $ 0.63 \pm 0.09 $ & $ 0.11 \pm 0.14 $ & $ 0.028 \pm 0.002 $ & $\phantom{*}8336.9 \pm 2375.3 $ & $ 0.60 \pm 0.04 $ \\
    \textbf{Deepali~\cite{deepali}} & $ 0.66 \pm 0.09 $ & $\underline{0.08 \pm 0.15}$ & $\underline{0.024 \pm 0.003}$ & $\phantom{*}6318.7 \pm 2330.4 $ & $ 0.24 \pm 0.03 $ \\
    \textbf{VXM~\cite{balakrishnan2019}} & $\underline{0.69 \pm 0.09}$ & $ 0.09 \pm 0.16 $ & $ 0.025 \pm 0.003 $ & $\phantom{*}\underline{6353.1 \pm 2328.4}$ & $ 0.25 \pm 0.02 $ \\
    \textbf{GAN~\cite{dey2021generative}} & $ 0.67 \pm 0.09 $ & $ 0.11 \pm 0.16 $ & $ 0.026 \pm 0.003 $ & $\phantom{*}6652.6 \pm 2303.7 $ & $ 0.21 \pm 0.02 $ \\
    \textbf{Cond. CNN~\cite{dalca2019learning}} & $ 0.65 \pm 0.09 $ & $ 0.09 \pm 0.16 $ & $\underline{0.024 \pm 0.003}$ & $\phantom{*}6417.3 \pm 2349.4 $ & $ \mathbf{0.15 \pm 0.02} $ \\
    % \midrule
    % \textbf{Img LDM} & $ 0.69 \pm 0.10 $ & $ 1.09 \pm 0.27 $ & $ 0.048 \pm 0.004 $ & $ 21944.0 \pm 2262.5 $ & $ 0.38 \pm 0.04 $\\
    % \textbf{W/o LDM} & -& -& -& -& -\\
    
    \midrule
    \textbf{DiffDeff [Ours]} & $ \mathbf{0.71 \pm 0.09} $ & $ \mathbf{0.06 \pm 0.15} $ & $ \mathbf{0.023 \pm 0.003} $ & $\mathbf{\phantom{*}5914.4 \pm 2289.2} $ & $\underline{0.19 \pm 0.02}$  \\
    \bottomrule
    % \multicolumn{6}{c}{ } \\
    % \multicolumn{6}{c}{ } \\
    \toprule
    \multicolumn{6}{c}{\textbf{Ventricular Volume}} \\
    \midrule
    & \textbf{DSC} $\uparrow$ & \textbf{Folding (\%)}  $\downarrow$ & \textbf{Smoothness} $\downarrow$ & \textbf{Avg. disp.} $\bm{\lVert \Phi \rVert}$ $\downarrow$ & \textbf{LPIPS} $\downarrow$\\ 
    \midrule
    \textbf{Linear} & $ 0.68 \pm 0.08 $ & $ 0.11 \pm 0.11 $ & $ 0.028 \pm 0.003 $ & $ \phantom{*}7782.9 \pm 2317.1 $ & $ 0.58 \pm 0.04 $ \\
    \textbf{Deepali~\cite{deepali}} & $ 0.70 \pm 0.08 $ & $\underline{0.07 \pm 0.12}$ & $\underline{0.025 \pm 0.003}$ & $\phantom{*}6408.1 \pm 2202.5 $ & $ 0.32 \pm 0.05 $ \\
    \textbf{VXM~\cite{balakrishnan2019}} & $ 0.69 \pm 0.08 $ & $ 0.09 \pm 0.12 $ & $ 0.026 \pm 0.003 $ & $\phantom{*}\underline{6016.0 \pm 2283.4}$ & $ 0.27 \pm 0.03 $ \\
    \textbf{GAN~\cite{dey2021generative}} & $\underline{0.71 \pm 0.07}$ & $ 0.12 \pm 0.14 $ & $ 0.026 \pm 0.003 $ & $\phantom{*}6767.0 \pm 2247.3 $ & $ 0.20 \pm 0.02 $ \\
     \textbf{Cond. CNN~\cite{dalca2019learning}} & $ 0.66 \pm 0.06 $ & $ 0.09 \pm 0.14 $ & $\underline{0.025 \pm 0.003}$ & $\phantom{*}6548.6 \pm 2223.6 $ & $\mathbf{0.16 \pm 0.02}$ \\
    %  \midrule
    % \textbf{Img LDM} & $ 0.68 \pm 0.13 $ & $ 0.98 \pm 0.23 $ & $ 0.047 \pm 0.003 $ & $ 19106.7 \pm 2113.5 $ & $ 0.37 \pm 0.03 $ \\
    % \textbf{W/o LDM} & $ 0.69 \pm 0.08 $ & $ \mathbf{0.05 \pm 0.12} $ & $ \mathbf{0.023 \pm 0.003} $ & $\phantom{*}\mathbf{5962.6 \pm 2110.6} $ & $ 0.20 \pm 0.02 $ \\
    \midrule
    \textbf{DiffDeff [Ours]} & $ \mathbf{0.75 \pm 0.07} $ & $ \mathbf{0.05 \pm 0.11} $ & $ \mathbf{0.023 \pm 0.003} $ & $\phantom{*}\mathbf{5354.1 \pm 2255.7} $ & $\underline{0.19 \pm 0.02} $\\
    
    \bottomrule
    \end{tabular}
}
\label{tab:results}
\end{table*}

\subsection{Evaluation} Evaluating the construction of an atlas poses a challenge, as ground truth is not available for comparison. The conditional atlas should ideally satisfy two competing criteria. 
Firstly, it should minimise the distance to every subject that satisfies the condition, and secondly, it should remain equally distant from all the subjects. 
As a result, we decided to evaluate our method along three key aspects that assess these desired properties: quantitatively assessing their structural properties, quantitatively measuring their appearance plausibility, and qualitatively assessing their visual appearance.

To perform the quantitative evaluation, we segment each generated atlas to obtain ventricle labels. 
We do so by segmenting~\cite{billot_synthseg_2023} the general population atlas and deforming the labels to each generated atlas. 
We then register a test set of \(100\) images that satisfy the condition onto the conditional atlas, resulting in a deformation field \(\phi_{i}\) for each image. 
We then assess the centrality of the atlas, \emph{i.e.} its distance from every subject in the test set, by comparing the average norms of the displacements (\(\frac{1}{100}\sum^{}_{i}\lVert\phi_{i}\rVert\)), the spatial smoothness with the the gradient magnitude of the transformations' Jacobian determinant (\(|\nabla_{J}|\)), and the foldings with the ratio of points with the percentage of points with a negative Jacobian determinant \(|J|<0\).
Furthermore, we assess the structural plausibility by reporting the mean and the standard deviation of the Dice overlap between the test set labels and the generated atlas.

% Finally, we take the average norm deformation field to calculate the atlas's ability to generate representative samples. 
To quantify the image quality, \emph{i.e.} whether the atlas appearance is similar to the test set, we compute the Learned Perceptual Image Patch Similarity (LPIPS) \cite{zhang2018unreasonable}.

\begin{figure}[ht!]
    \centering
    \includegraphics[width=\columnwidth]{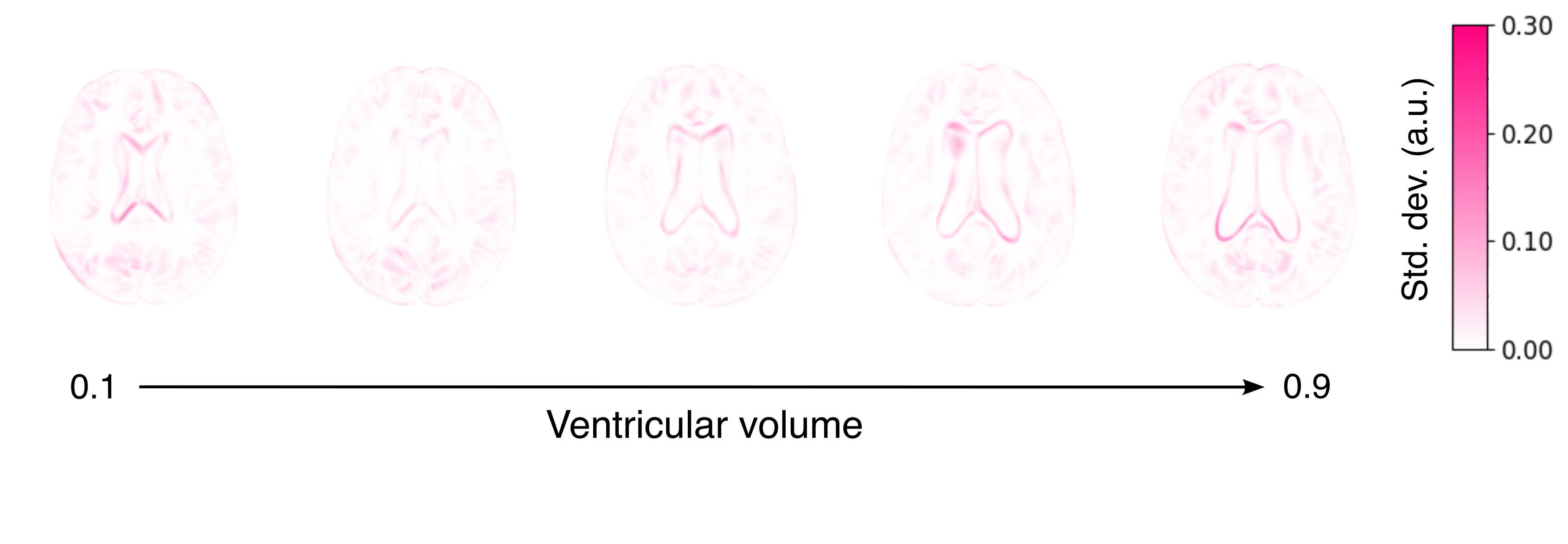}
    \caption{Visualisation of variance across three atlases sampled using our method with three different noise patterns, shown for increasing ventricular volume. 
    The results demonstrate the model’s ability to produce anatomically plausible atlases with minimal variance.
    The colourbar denotes the standard deviation (a.u.).}
    \label{fig:variance_maps}
\end{figure}

\begin{figure*}
    \centering
    \includegraphics[width=0.9\textwidth]{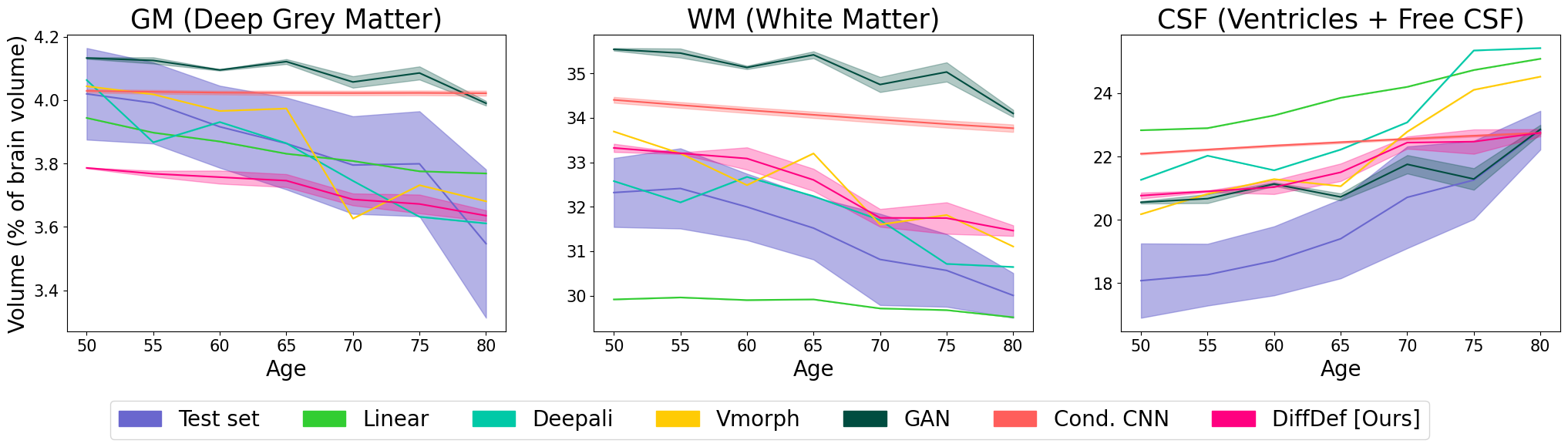}
    \caption{Segmentation label volume percentages of atlases conditioned on age. 
    Our method accurately captures age-related anatomical changes, including grey and white matter atrophy and increased cerebrospinal fluid volume with growing age.
    For the learning-based models, i.e. Ours, Cond. CNN and GAN, we report the mean and standard deviation obtained by sampling across three different random seeds. The test set consists of 100 subjects, with the standard deviation capturing the variability within this cohort. It serves as a baseline, reflecting the natural trends present in the data.
    }
    \label{fig:quant_age}
\end{figure*}

\begin{figure*}
    \centering
    \includegraphics[width=\textwidth]{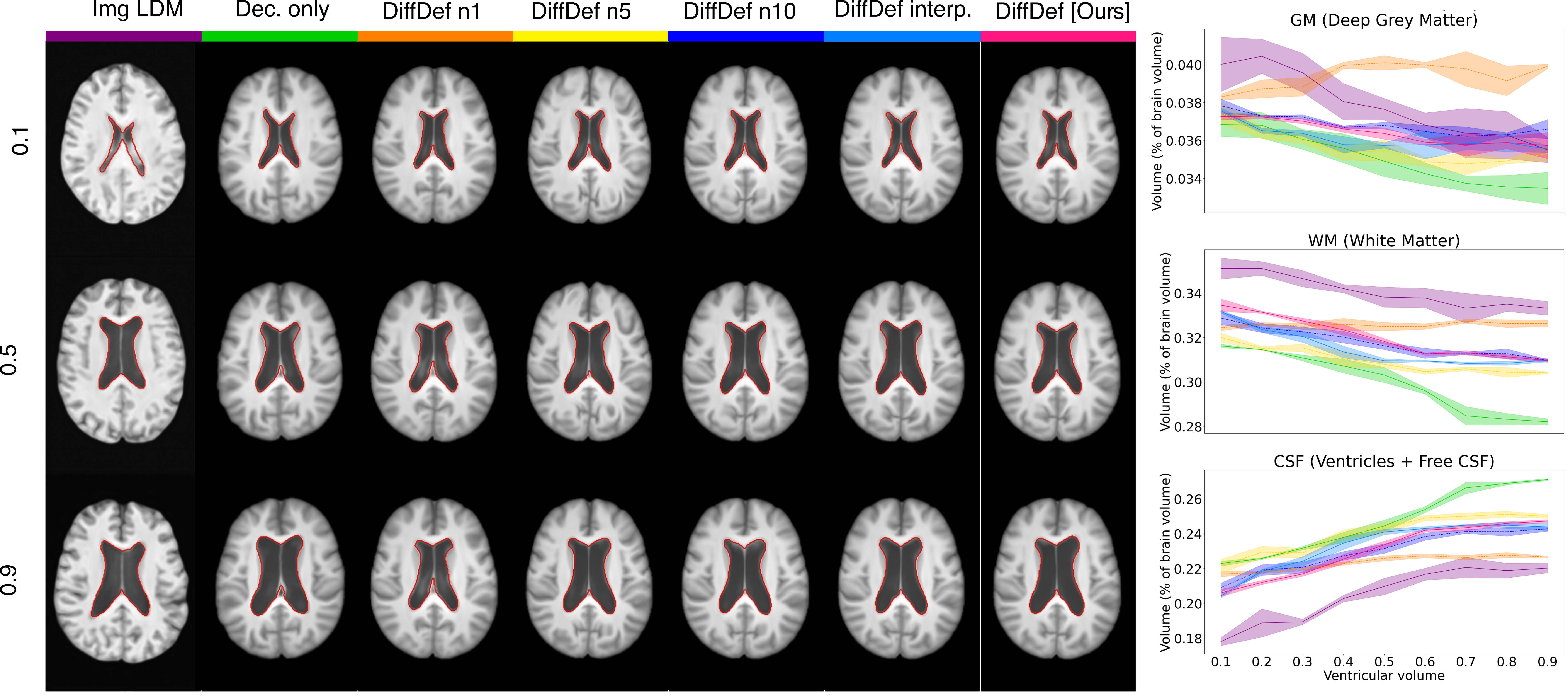}
    \caption{Qualitative results of all ablated experiments across increasing normalised ventricular volume sizes (\(0.1,0.5,0.9\)).
    The ventricles are delineated in red to enhance visualisation and emphasise the changes in ventricular size. 
    On the right, we present the mean and standard deviation segmentation label volume percentages of atlases generated by each baseline, conditioned on ventricular volume.}
    \label{fig:neighborhood-ablation}
\end{figure*}

\section{Results and Discussion}

The optimal conditional atlas should minimise the distance to all subjects that satisfy the query condition.
In the following, we evaluate (a) the qualitative results of our method and the comparable baseline results (see Figures~\ref{fig:csf} and \ref{fig:qualitative}), (b) the quantitative results by using metrics that quantify similarities in appearance, structural properties, and centrality (listed in Table~\ref{tab:results} and Figure~\ref{fig:quant_age}) and (c) the generalisability potential of our proposed approach (Figure~\ref{fig:neighborhood-ablation} and Table~\ref{tab:ablation_results}).

Figure~\ref{fig:qualitative} illustrates the resulting brain atlases of all the different methods conditioned on the ventricular volume and age.
% While the ventricular volume does not have a clear clinical relevance, we decided to include results conditioned on it because the associated changes in image appearance are more visually apparent, making them easier to interpret qualitatively.
Comparing our method (last column) to the conventionally generated atlases (Linear, Deepali and VXM), we achieve sharper boundaries while maintaining the intensity distribution of the dataset and the accurate morphological features.
Since our method deforms an existing population atlas with the generated deformation field, it does not introduce any intensity shift. 
Moreover, the deformation field is regularised during training, ensuring that no unrealistic or out-of-distribution anatomical structures are generated.
In contrast, GAN and Cond. CNN are prone to generate unrealistic intensities and noisy backgrounds as acknowledged in~\cite{dey2021generative}. They require masking as an additional post-processing step to mitigate this effect. 
Furthermore, brain shapes for both GAN and Cond. CNN vary noticeably from the \emph{expected} brain shape that the conventional methods compute.
Indeed, the frontal lobe is narrower both for the GAN and Cond. CNN-generated case, in all generated conditions.

An increase in ventricular volume due to the atrophy of the surrounding brain tissue is a well-studied biomarker in neurological ageing~\cite{kaye1992significance}. This is visible in all three conventional methods, GAN and our approach, while Cond. CNN fails to capture this effect consistently.
Furthermore, our approach is the only one that generates a deformation field. 
This inherently enhances the interpretability of our method, allowing us to localise structural changes. 
We illustrate this in  Figure~\ref{fig:csf}, where the Jacobian determinants of the generated displacements are visualised alongside the final produced atlases conditioned on ventricular volume. 
The Eigenvalues of the Jacobian determinant indicate the magnitude of expansion (red) or compression (blue) in the image domain. 

\begin{table*}[h!]
\caption{Overall efficiency regarding runtime and sample usage for atlas creation.
Generative methods require longer training, enable fast inference without additional samples, and can interpolate missing conditions. 
In contrast, non-generative methods need many subjects per condition and involve time-consuming registration during atlas generation.
We use Atlas creation samples to indicate whether a method requires condition-specific subjects for atlas generation.}
\label{tab:speed}
% \addtolength{\tabcolsep}{0.5pt} % Adjust column padding if needed
\resizebox{\textwidth}{!}{
\centering{
\begin{tabular}{lcccccc}
\midrule
\textbf{Method} & \textbf{Linear} & \textbf{Deepali} & \textbf{VXM} & \textbf{GAN} & \textbf{Cond. CNN} & \textbf{DiffDef [Ours]} \\
\toprule
% \textbf{Training samples} & N/A& N/A& $5000$ & $5000$& $5000$ & $5000$\\
\textbf{Atlas creation samples} & Yes & Yes & Yes & No & No & No\\
\textbf{Training time} & N/A & N/A & $12$ hours & $5$ days & $5$ days & $1$ day \\
\textbf{Atlas generation time} & $297 \pm 15$s & $2782 \pm 57$ & $132.26 \pm 5.15$s & $1.12 \pm 0.35$s & $0.58 \pm 0.37$s & $24.63  \pm 0.13$s\\
\bottomrule
\end{tabular}
}
}
% }
\end{table*}

To quantitatively evaluate our results, we select a test set of $100$ images per condition, which we register to each conditional atlas.
% We then assess the structural plausibility by measuring the average Dice overlap, the amount of folding and the spatial smoothness of transformation.
To evaluate the structural plausibility, we segment the population atlas using SynthSeg~\cite{billot_synthseg_2023} to obtain the ventricle labels, which we propagate to the generated atlases via deformable registration.
Then, we measure the Dice overlap of the conditional atlases with each test set image.  
Additionally, we evaluate the spatial folding, reporting the percentage of points with \(J<0\) and the smoothness with the magnitude of the gradient of the Jacobian determinant (\(|\nabla_{J}|\)). 
Finally, to evaluate whether the generated atlases deviate in appearance from real images in the test set, we employ the Perceptual Image Patch Similarity (LPIPS) metric~\cite{zhang2018unreasonable}.
In Table~\ref{tab:results}, we demonstrate that the proposed method, DiffDeff, demonstrates superior performance in all metrics compared to the conventional methods. 
In particular, it improves the structural similarity indicated by the meqn Dice score by \(2\%\) in the age case and \(4\%\) in the ventricular volume case while also being spatially smoother, demonstrating a lower folding ratio and smoothness metric.
Finally, the generated atlases exhibit the lowest centrality for both conditioning scenarios, measured by the average deformation norm, indicating that our method yields the most representative atlases. 
While our method ranks second in terms of perceptual similarity, this indicates that the Cond. CNN produces atlases that more closely resemble the test set in appearance. 
However, as illustrated in Figure~\ref{fig:qualitative}, the atlases generated by the Cond. CNN are of lower anatomical quality. 
This suggests that, although they may visually resemble individual test samples, they lack the generalisability and representativeness expected of a population-level atlas.

A key advantage of the proposed method is its ability to generate fast, accurate, and robust atlases. 
As shown in Figure~\ref{fig:quant_age}, our model successfully captures established ageing-related trends reported in the literature~\cite{smith2007age}, including the shrinkage of grey and white matter and the corresponding increase in cerebrospinal fluid volume with advancing age. In contrast, generative models such as GAN fail to capture meaningful anatomical trends.
Additionally, Figure~\ref{fig:csf} highlights that our model is also able to capture ventricular volume growth, which is frequently associated with neurodegenerative disorders and has been linked to cognitive decline~\cite{nestor2008ventricular,ott2010brain,crook2020linear}.   

Moreover, as illustrated in Figure~\ref{fig:variance_maps}, our model exhibits robust performance, consistently generating atlases with minimal variance across three independent samplings using different noise patterns. 
This robustness is especially evident across varying ventricular volume sizes, highlighting the proposed model’s ability to produce anatomically plausible atlases with high stability and low variability.

Lastly, Table~\ref{tab:speed} summarizes the training and atlas generation speeds of each method, along with the sample requirements during atlas construction. 
Generative methods offer substantial improvements in speed over conventional approaches (i.e., Linear, Deepali, and Voxelmorph), achieving up to a 
\(99.96\%\) reduction in processing time. 
An additional benefit of these generative approaches is their independence from extra samples during atlas generation once training is complete. 
This enables the use of the entire dataset for model training, thereby enhancing the ability to capture the underlying data distribution. 
In contrast, conventional methods demand a significant number of subjects for each condition, which can be impractical, especially for underrepresented groups such as specific age ranges.
Another noteworthy observation is that, while our method remains significantly faster than conventional approaches, it is comparatively slower than other generative methods, specifically GAN and Conditional CNN—when generating conditional atlases. 
This is primarily due to using a diffusion model as the backbone, which requires multiple steps to produce a conditional atlas. 
However, our method compensates for this with a notably faster training time, approximately five times faster, making it far more practical for tuning and deployment. In contrast, training the GAN and Conditional CNN baselines proved to be both time-consuming and technically demanding. 
These models frequently encountered training instability during hyperparameter tuning, including issues such as mode collapse. 
Overcoming these challenges required substantial data curation and a heavy reliance on checkpointing. 
In our assessment, these limitations significantly reduce the practical viability of these approaches.

\begin{table*}[h]
% \small
\addtolength{\tabcolsep}{2pt}
\centering
    \caption{Quantitative results that assess competing properties (anatomical and appearance similarity) of the generated atlases of the ablation experiments conditioned on ventricular volume. 
    We denote models that use a diffusion backbone as LDM, models that generate deformation fields instead of image intensities as \(\phi\), and models trained with the complete set of conditions as All. Cond. The term \# N. indicates the number of neighbours used during training.
    We perform pairwise comparisons between each generated atlas and a test set of $100$ images for each condition and report each metric's mean and standard deviation.
    The best results are highlighted in \textbf{bold}, and the second best are \underline{underlined}.}
    \resizebox{\textwidth}{!}{
    \begin{tabular}{lccccccccc}

    \toprule
    \multicolumn{10}{c}{\textbf{Ventricular Volume}} \\
    \midrule
        & \textbf{LDM} & $\mathbf{\phi}$ & \textbf{All Cond.} & \textbf{\# N} &\textbf{DSC} $\uparrow$ & \textbf{Folding (\%)}  $\downarrow$ & \textbf{Smoothness} $\downarrow$ & \textbf{Avg. disp.} $\bm{\lVert \Phi \rVert}$ $\downarrow$ & \textbf{LPIPS} $\downarrow$\\ 
    \midrule

    \textbf{Img LDM} & \ding{51} & \ding{55} & \ding{51} & 15 & $ 0.682 \pm 0.129 $ & $ 0.975 \pm 0.227 $ & $ 0.047 \pm 0.003 $ & $ 19106.7 \pm 2113.5 $ & $ 0.372 \pm 0.030 $\\
    \textbf{Dec. only} & \ding{55} & \ding{51} & \ding{51} & 15 & $ 0.690 \pm 0.081 $ & $ 0.052 \pm 0.122 $ & $ \mathbf{0.023 \pm 0.003} $ & $ \phantom{*}5962.6 \pm 2110.6 $ & $ 0.197 \pm 0.024 $ \\
    \midrule
    \textbf{DiffDef n1} & \ding{51} & \ding{51} & \ding{51} & 1 & $ 0.720 \pm 0.065 $ & $ 0.048 \pm 0.109 $ & $ \mathbf{0.023 \pm 0.003} $ & $ \phantom{*}5382.9 \pm 2179.6 $ & $ 0.193 \pm 0.023 $ \\
    \textbf{DiffDef n5} & \ding{51} & \ding{51} & \ding{51} & 5 & $ 0.749 \pm 0.068 $ & $ 0.057 \pm 0.112 $ & $ \underline{0.024 \pm 0.003} $ & $ \phantom{*}\mathbf{5203.0 \pm 2206.8} $ & $ 0.189 \pm 0.021 $ \\
    \textbf{DiffDef n10} & \ding{51} & \ding{51} & \ding{51} & 10 & $ \underline{0.750 \pm 0.069} $ & $ 0.047 \pm 0.113 $ & $ \mathbf{0.023 \pm 0.003} $ & $ \phantom{*}\underline{5300.3 \pm 2275.2} $ & $ \underline{0.191 \pm 0.022} $ \\
    \textbf{DiffDef interp.} & \ding{51} & \ding{51} & \ding{55} & 15  & $ \underline{0.750 \pm 0.072} $ & $ \mathbf{0.044 \pm 0.112} $ & $ \mathbf{0.023 \pm 0.003} $ & $ \phantom{*}5361.3 \pm 2257.7 $ & $ \mathbf{0.190 \pm 0.023} $ \\

    \midrule
    \textbf{DiffDef [Ours]} & \ding{51} & \ding{51} & \ding{51} & $15$ & $ \mathbf{0.755 \pm 0.067} $ & $ \underline{0.045 \pm 0.114} $ & $ \mathbf{0.023 \pm 0.003} $ & $\phantom{*}5354.1 \pm 2255.7 $ & $ \mathbf{0.190 \pm 0.023} $\\

    \bottomrule
    \end{tabular}
}
\label{tab:ablation_results}
\end{table*}

\begin{figure}
    \centering
    \includegraphics[width=\columnwidth]{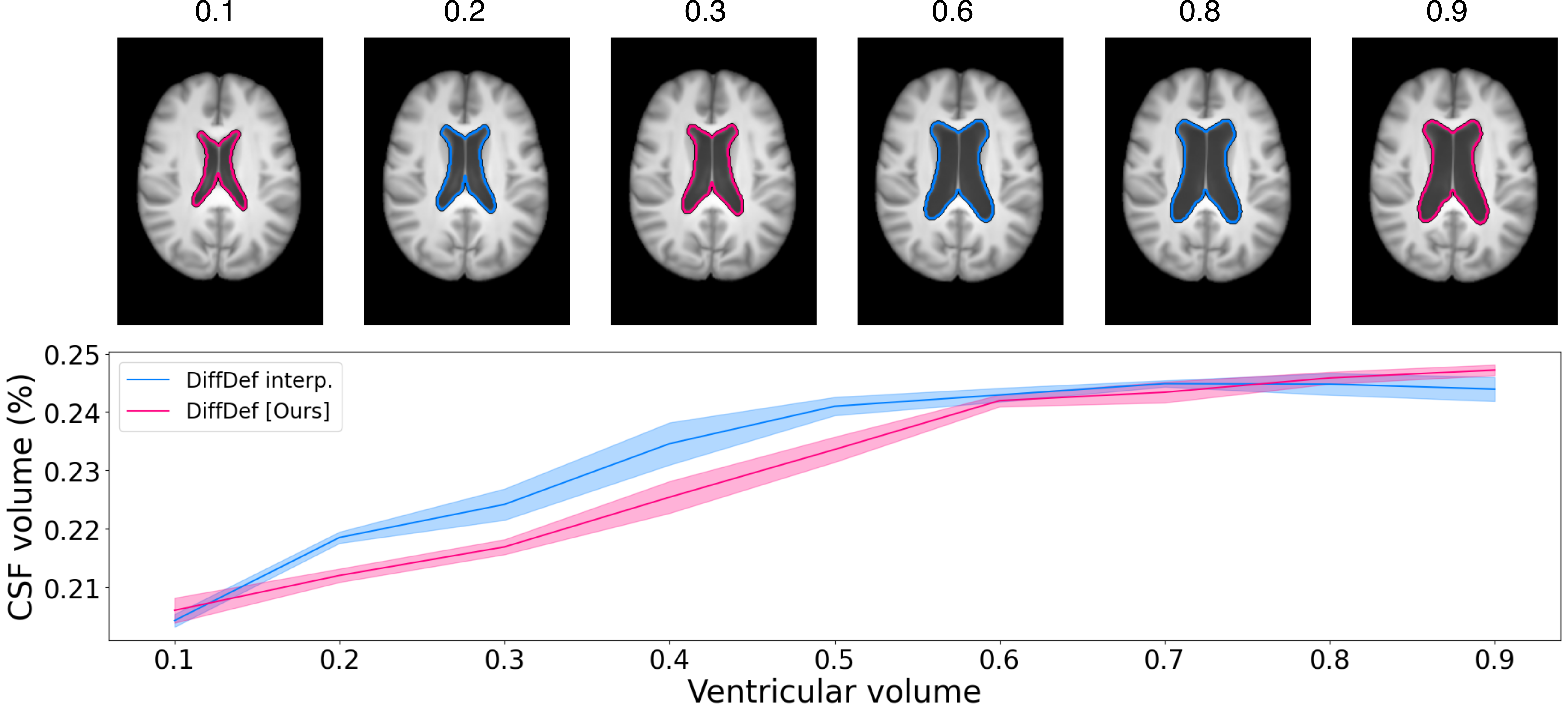}
    \caption{Evaluation of the model's generalisation to unseen conditioning values. The model is trained using a subset of ventricular volume levels (0.1, 0.3, 0.5, 0.7, and 0.9) and used to generate atlases for both the training conditions and the intermediate, unseen conditions (0.2, 0.4, and 0.8). The ablated model (\textcolor{azureblue}{blue}) successfully generalises to unseen conditions, while the full model (\textcolor{hotpink}{pink}) produces similar CSF volume quantification. For both models, we report the mean and standard deviation obtained by sampling across three different random seeds.}
    \label{fig:abl_interp}
\end{figure}

\subsection{Ablations}

We perform a series of ablation studies, summarised in Table~\ref{tab:ablation_results}, to assess each component's contribution to the proposed framework, targeting (1) the impact of incorporating the diffusion model, (2) the influence of the neighbourhood size, (3) the effect of generating deformation fields versus directly synthesising atlas intensities, and (4) the model’s ability to generalise to unseen conditioning attributes.
For a qualitative inspection of the atlases generated for the ablation experiments, along with the volume percentages of grey and white matter and cerebrospinal fluid, we refer to Figure~\ref{fig:neighborhood-ablation}.

First, to assess the contribution of the diffusion model in conditional atlas synthesis, we eliminate both the latent diffusion module and the autoencoder. 
In their place, we implement a convolutional decoder that directly maps the conditioning variable to a deformation field. 
This simplified baseline, referred to as Decoder only (Dec. only) in Table~\ref{tab:ablation_results}, is trained using the morphology preservation loss described in Section~\ref{sec:morphology_preservation}.
This architectural modification offers certain advantages, including reduced memory usage and computational complexity due to the lightweight decoder replacing the LDM.
Nevertheless, as shown in Table~\ref{tab:ablation_results}, our model with the LDM backbone achieves a significantly higher mean Dice score of $0.75$ than the Decoder-only baseline $0.69$. 
We attribute this improvement to the superior capacity of the latent diffusion model to capture the underlying data distribution.
The randomly sampled latent vector used in the Dec. only experiment lacks the structured information necessary for the decoder to generate coherent, high-quality data. Without guidance, the decoder must learn an extremely complex mapping from unstructured noise to structured outputs, often resulting in poor sample quality or unstable training.
The progressive refinement of the proposed method allows the model to explore the data distribution in a controlled, structured manner, capturing subtle details and complex correlations that would be difficult for a simple decoder to learn. 
The denoising process effectively guides the model through the generative pathway, resulting in more stable, expressive, and high-fidelity outputs.

Second, to investigate the impact of generating deformation fields versus directly predicting atlas intensities, we retain the latent diffusion model but modify its output to produce atlas intensities instead of deformation fields. 
As in the previous setup, training is guided by the proposed morphology preservation loss (Section~\ref{sec:morphology_preservation}).
We denote this experiment in Table~\ref{tab:ablation_results} as Img LDM.
However, this configuration performs poorly in practice, as seen quantitatively in all metrics demonstrated in Table~\ref{tab:ablation_results} and qualitatively in Figure~\ref{fig:neighborhood-ablation}. 
We attribute this to the inherent difficulty of intensity generation, which requires precise pixel-wise correspondence across subjects with varying anatomies and acquisition conditions. 
This challenge often results in unstable training dynamics or collapsed outputs.
We hypothesise that prior works, such as those by Dalca et al.~\cite{dalca2019learning} and Dey et al.~\cite{dey2021generative}, mitigate this difficulty by generating only residual intensities, which are added to a linearly constructed atlas, thereby reducing the complexity of the learning task. 
In contrast, generating deformation fields directly offers a more robust and anatomically meaningful approach to conditional atlas synthesis. 
By modelling geometric transformations instead of raw intensities, this strategy circumvents issues related to pixel-level alignment and inter-subject intensity variability.

Third, we investigate the effect of neighbourhood size in the conditioning set by experimenting with $n=1,5,10,15$. 
These experiments aim to demonstrate that a larger neighbourhood contributes to a more stable and consistent atlas generation.
This trend is evident in Table~\ref{tab:ablation_results}, where larger neighbourhood sizes result in higher Dice overlap and improved centrality metrics. 
In this work, we select a neighbourhood size of 15, as it represents the largest configuration that fits within our available GPU memory constraints.

Finally, to evaluate the model's ability to generalise to unseen conditioning values, we train the model using a subset of ventricular volume levels, specifically 0.1, 0.3, 0.5, 0.7, and 0.9, and generate atlases for both these training values and the unseen intermediate conditions 0.2, 0.4, and 0.8.
We denote this model in Table~\ref{tab:ablation_results} as DiffDeff interp.
Notably, the model architecture, hyperparameters and training procedure remained unchanged; only the training input data was modified by excluding data whose ventricular volume size matches these specific conditions.
The metrics presented in Table~\ref{tab:ablation_results} and Figurse~\ref{fig:neighborhood-ablation} and~\ref{fig:abl_interp} demonstrate our method's robustness in managing missing conditioning values. 
This capability allows the model to infer these conditions exclusively from the learned distribution, which is particularly advantageous for addressing challenges associated with unbalanced or incomplete datasets.

\section{Conclusion}
Atlases generated with conventional methods are well-established due to their reliability and realism. 
They, however, face scalability issues in terms of speed, data, and memory requirements, which renders them difficult to use with sub-population conditioning. 
To address this, generative modelling has been used to synthesise conditional atlases, which is faster and not as dependent on the conditioning variable. 
However, this comes with other limitations, such as training instabilities and mode collapse. 
In this work, we propose to combine the highly interpretable deformation vector field from conventional methods and the power of diffusion models to \emph{generate deformation fields} that transform an existing population atlas into conditioned ones. 

We train a conditional latent diffusion model to generate deformation vector fields, which transforms a general atlas into a conditional one to match the query condition. 
We jointly train a morphology-preserving network that enforces the conditioning feature to be satisfied with respect to a neighbourhood.
Our proposed method outperforms previous approaches in terms of structural and perceptual aspects. 
Moreover, it is able to generalise at inference to conditions unseen during training. 
While our approach shares the high resource demands typical of generative methods, it remains comparatively more efficient in terms of training time and computational cost. 
However, it is dependent on the availability of a population atlas. 
In the case of the brain, this is a minor limitation, as several high-quality atlases already exist. 
In contrast, defining and constructing atlases for other image types, such as whole-body scans, is more challenging due to greater anatomical variability. 
As a future direction, extending our analysis to include demographic and pathology-related attributes beyond age could offer deeper insights into condition-specific brain changes.
Conversely, once trained, it can generate conditional atlases in seconds. 
Finally, it is not tailored to a specific image modality; one could learn to generate an atlas on a T1-weighted dataset and seamlessly extend it to another modality.

\bibliographystyle{IEEEtran}
\bibliography{IEEEabrv,literature}

\end{document}